\documentclass[11pt,a4paper]{article}
\usepackage{jheppub,amsmath,  amssymb,slashed,url,bm,textgreek,upgreek}
\usepackage{graphicx}
\usepackage{epstopdf}
 
\def\tt{{\mathfrak t}}
\def\J{\mathcal J}

\def\RR{{\mathcal R}}

\def\be{\begin{equation}}
\def\ee{\end{equation}}

\def\eff{{\mathrm{eff}}}
\def\tilde{\widetilde}

\def\Bbb{\mathbb}

\def\d{{\mathrm d}}

\def\intt{{\matheurm{int}}}

\def\R{{\mathbb R}}

\def\[{\bigl [}

\def\]{\bigr ]}

\def\Z{{\mathbb Z}}

\def\L{{\mathcal  L}}

\def\I{{\mathcal I}}

\def\M{{\mathcal M}}

\def\tilde{\widetilde}

\font\teneurm=eurm10 \font\seveneurm=eurm7  \font\fiveeurm=eurm5
\newfam\eurmfam
\textfont\eurmfam=\teneurm \scriptfont\eurmfam=\seveneurm
\scriptscriptfont\eurmfam=\fiveeurm

\font\teneusm=eusm10 \font\seveneusm=eusm7 \font\fiveeusm=eusm5
\newfam\eusmfam
\textfont\eusmfam=\teneusm \scriptfont\eusmfam=\seveneusm
\scriptscriptfont\eusmfam=\fiveeusm

\font\tencmmib=cmmib10 \skewchar\tencmmib='177
\font\sevencmmib=cmmib7 \skewchar\sevencmmib='177
\font\fivecmmib=cmmib5 \skewchar\fivecmmib='177
\newfam\cmmibfam
\textfont\cmmibfam=\tencmmib \scriptfont\cmmibfam=\sevencmmib
\scriptscriptfont\cmmibfam=\fivecmmib

\def\Maxwell{{\rm Maxwell}}
\def\Tr{{\rm Tr}}
\def\CS{{\rm CS}}
\def\Z{{\Bbb Z}}
\def\R{{\Bbb R}}
\def\d{{\mathrm d}}
\def\eff{{\rm eff}}
\def\micro{{\rm micro}}
\def\tilde{\widetilde}
\def\J{{\mathcal J}}
\def\i{{\mathrm i}}
\def\intt{{\rm int}}

\title{The Chern-Simons Function and the Quantum Hall Effect}

 \author{Edward Witten}
\affiliation{School of Natural Sciences, Institute for Advanced Study,\\ 1 Einstein Drive, Princeton, NJ 08540 USA}
\abstract{In this article,\footnote{The article is based on a lecture presented at a conference celebrating the 75th birthday of James Simons, held  in 2013 at the CUNY Graduate Center.   The article will
appear  in an issue of the Bulletin of the American Mathematical Society in memory of Simons.} I will define the Chern-Simons invariant of a gauge connection over a three-manifold and explain some of the things that
are special about it, in particular its role in understanding the quantum Hall effect, which is an important effect in condensed matter physics,
and its generalization, the fractional quantum Hall effect.  
}

\begin{document}\maketitle

\section{The Chern-Simons Function Of A Connection}\label{one}

Instantons in Yang-Mills theory \cite{BPST} became important in physics in the mid-1970's when it was found \cite{THooft} that they provide the key to resolving an apparent contradiction between experimental data and the then emerging standard model of particle physics.   From a mathematical point
of view, an instanton is a connection on a vector bundle $V$ over a smooth oriented four-manifold $Y$, with structure group a compact simple Lie group $G$, that obeys a certain nonlinear differential equation.  
Later, it turned out that  the instanton equation is a powerful tool for studying smooth four-manifolds \cite{Donaldson}.

An  instanton bundle $V\to Y$ can be  characterized  topologically by an integer,
the integral over $Y$ of a certain characteristic class.  This integer is known as the instanton number.
 For example, if $G=SU(N)$, the relevant characteristic class is the second Chern class, and the instanton number can be defined by an integral
\be\label{firstint}\I_Y=\int_Y\, \frac{\Tr\,F\wedge F}{8\pi^2}, \ee
where $F$ is the curvature of a connection $A$ and the trace is taken in the $N$-dimensional representation.
This quantity is integer-valued for any $SU(N)$ bundle $V\to Y$, and if $V$ admits a connection that satisfies the instanton equation, then the integer is non-negative.

Before the 1970's, topology in general and the second Chern class in particular were  certainly not on the radar screen of most theoretical physicists.
In large part, instantons and their role in  the standard model introduced physicists to the importance  of topology in gauge theory and in quantum theory more generally.

Related to the second Chern class in four dimensions is a ``secondary characteristic class'' in three dimensions, namely the Chern-Simons function.
Let $W$ be an oriented three-manifold and $V\to W$ a $G$-bundle with connection $A$. If $W$ is the boundary of an oriented four-manifold $X$ over which $V$ and $A$ can be extended, which is always the case for $G=SU(N)$, then the Chern-Simons function can be
defined\footnote{Physicists usually include an extra factor of $2\pi$ in this definition and consider $\CS(A)$ to be valued in $\R/2\pi\Z$.}  by the same   integral over $X$
that gives the second Chern class if the boundary of $X$ is empty:
\be\label{csdef}\CS_X(A)=\int_X\, \frac{\Tr\,F\wedge F}{8\pi^2}. \ee
Because $X$ has a boundary, this integral is not a topological invariant, and it does depend on the data on $X$, not just on $W$.  However, mod $\Z$, it depends only on the restriction of $A$ to the boundary $W$.  To prove this, consider
two extensions of $V$ and $A$ over possibly different oriented four-manifolds $X$ and $X'$, each with the same boundary $W$.   By gluing together $X$ and $X'$ along their common
boundary, after reversing the orientation of $X'$, one builds a closed oriented four-manifold $Y$ endowed with an $SU(N)$ bundle $V\to Y$, and one has
\be\label{nsdef}\CS_X(A)-\CS_{X'}(A)=\I_Y\in \Z.\ee
Since $\CS_X(A)$ is thus independent of $X$ mod $\Z$, we can forget $X$ if we consider $\CS(A)$ to be valued in $\R/\Z$.   This $\R/\Z$-valued function of a connection or gauge field
on an oriented  three-manifold 
is the Chern-Simons function or the Chern-Simons invariant of the connection.

Once it entered physics via its role in instanton theory, the Chern-Simons function turned out to have many applications, too many to summarize in this article.   
In this article,  I will explain just one part of the story, which is the role of the Chern-Simons function in understanding an effect in condensed matter physics that is known as the quantum Hall effect.
In fact, we will begin with the more simple integer quantum Hall effect, and then more briefly describe  the  fractional case.

One bit of physics that I will assume is familiar is that in vacuum, light is described by Maxwell's equations.   In general relativity, spacetime is a four-dimensional manifold $M$
with a pseudo-Riemannian metric, that is a metric of signature $-+++$.   The canonical example (and a good idealization for many purposes) is Minkowski 
 space
$\R^{1,3}$, that is,  $\R^4$ with the pseudo-Riemannian metric
\be\label{turkey}\d s^2=-\d t^2+\d\vec x^2,\ee
where $\vec x=(x,y,z)$. 
   Maxwell's equations are equations for a two-form $F$, known as the electromagnetic field strength tensor.   The equations read
\be\label{delfo} \d F=\d \star F=0,\ee where $\star$ is the Hodge star operator on two-forms.   In Euclidean signature, one would describe
these equations by saying that $F$ is a harmonic two-form, a fact that may have been part of the inspiration for Hodge theory.

However, in the context of quantum mechanics, and for other reasons that we will see,
 instead of merely viewing $F$ as a two-form, it is better to interpret it as the curvature $F=\d A$
of  a connection or abelian gauge field $A$ on
a complex line bundle over $M$.   Then one Maxwell equation $\d F=0$ becomes an identity -- the Bianchi identity -- and the second one
$\d\star F=0$ is the Euler-Lagrange equation derived from the ``Maxwell action''
\be\label{nurkey} I_\Maxwell=\frac{1}{4 e^2}\int_{M }F\wedge \star F, \ee where $e$ is a constant, the charge of the electron. 
To verify this, we note that if we vary $A$ by $A\to A+\delta A$, where $\delta A$ is a small variation of $A$ that we will treat to first order, then
the curvature $F$ changes by $F\to F+\d\delta A$.  In first order, the change in $I$ is then
\be\label{lurkey}\delta I =\frac{1}{2e^2}\int_{M}\d\delta A\wedge \star F=\frac{1}{2e^2}\int_{M}\delta A\wedge\d \star F. \ee
So the condition $\delta I=0$ for $I$ to be stationary under variations of $A$ indeed gives the second Maxwell equation $\d\star F=0$.

We included in defining $I_\Maxwell$ a factor $1/e^2$ that is important quantum mechanically, but  plays no role whatsoever in classical physics.
In classical physics, the sole importance of the action is that the equations of motion are the Euler-Lagrange equations, which say that the action is stationary
under small variations.   That condition is invariant under multiplying the action by a constant so an overall multiplicative constant in the action is unimportant.
Indeed, Maxwell's equations do not depend on $e$.

\section{Insulators}\label{two}

In this article, we are going to consider light, and more general electromagnetic fields, interacting with a material.  But to keep things simple,
we are going to take the material to be an insulator, such as  a piece of glass.  An insulator for us is an inert substance,
which has no relevant degrees of freedom that we have to be concerned with.\footnote{At the end of the article, we will slightly relax this assumption
and assume merely that the insulator has no relevant {\it local} degrees of freedom.}  (Why such a material is an electrical insulator will be explained later.)
  So describing how light behaves in an insulator just means
describing how Maxwell's equations are modified in the presence of the insulator.  By contrast, to describe light interacting with a conducting material,
we would need to develop a theory of the charges and currents inside the material; it would be necessary to consider much more physics.

In our general discussion of insulators -- until we get to  two-dimensional materials -- nothing essential is lost if we simply take spacetime to be Minkowski
space $\R^{1,3}$.   Moreover, we assume that the insulator is at rest in some Lorentz frame, with time coordinate $t$ and spatial coordinates $\vec x$.  The region
$S\subset \R^{1,3}\cong \R\times \R^3$ that the insulator occupies is thus of the form $\R\times S_0$ for some region $S_0\subset \R^3$.
Given these assumptions, a nonrelativistic notation is natural.
We split  the two-form $F$ into the ``electric field'' 
 $\vec E$ and the ``magnetic field''  $\vec B$:
\be\label{indo} F=\d t\,\d\vec x\cdot \vec E +\frac{1}{2}\d\vec x\times \d\vec x \cdot \vec B. \ee  (Here $\cdot$ is the dot product of
two vectors in three-dimensional space, and $\times$ is the cross product.)   In this nonrelativistic notation, the Maxwell action 
can be rewritten as
\be\label{nindo}I_\Maxwell=\frac{1}{2e^2}\int\d t\,\d^3 x \left(\vec E^2-\vec B^2\right). \ee

It is not a good idea to formulate the problem of describing the interaction of light with the insulator directly in terms of modifying Maxwell's equations.
A random modification of Maxwell's equations would violate physical principles such as conservation of energy and conservation of momentum.  It is better
to modify the Maxwell action, not the equations; physically sensible modifications of Maxwell's equations are the ones that come from 
modifications of the action.     This is actually one of the reasons that it was a good idea to interpret the Maxwell field strength tensor $F$ as the curvature
of an abelian gauge field $A$.   This makes it possible to interpret Maxwell's equations in vacuum as the Euler-Lagrange equations associated to some action,
and thereby makes it possible to formulate the statement that the interactions of light with a material should be described by using a more general action.
The full action including the Maxwell action in vacuum and additional contributions due to the material make what we call
the ``effective action,'' one of the most important concepts in physics.

What modifications of the Maxwell action should we consider?  This depends on the material.   There are a lot of possible
materials, and  new ones are constantly being fabricated.  So more or less anything  we can imagine that is consistent with general principles can be realized by some material.
More specifically, we may consider  adding to the action $I$ the integral over $S$ 
of any polynomial in the electric and magnetic fields $\vec E $ and $\vec B$ and their derivatives..    Thus the effective action
may have a contribution
\be\label{tellme}I_\eff =\cdots +\int_S \d t\, \d^3x \, f(\vec E,\vec B,\partial \vec E,\partial \vec B,\partial^2\vec E,\partial^2\vec B,\cdots)\ee
with more or less any function $f$.  We assume that the material is homogeneous under space and time translations (away from its boundaries\footnote{The effective action can also include contributions supported on the boundary of $S$, and this can be important,
but we will not consider that possibility.}), or that if there are inhomogeneities (due to a crystal structure, for example), they are on too small a length scale for our experiments
to resolve.  So $f$ does not
depend explicitly on $t$ and $\vec x$.     In practice, experiments with macroscopic electromagnetic fields or with visible light will not resolve atomic scale inhomogeneities;
X rays are another matter.

Two things bring some order to this chaos.  First, for most purposes a Taylor series expansion of $f$ in powers of $E,B$, and their derivatives is very effective, with only a few terms being 
important.  That is true
because (i) the electric and magnetic fields that we can create in practice in the laboratory are small by atomic standards, so $\vec E$ and $\vec B$ can
be considered very small, and (ii) spacetime derivatives  can also be considered small for many purposes, since the electric and magnetic fields of interest are often slowly varying 
on an atomic scale.  This second approximation will fail if we consider the interaction of matter with X-rays or gamma rays, but it is an excellent approximation 
in the case of visible light and even better for the macroscopic electric and magnetic fields that are most relevant in the present article.

So we can think of the expansion in eqn. (\ref{tellme}) as a rapidly convergent expansion in $\vec E$, $\vec B$, and their derivatives.
The second thing that brings some order to chaos is that this expansion is constrained by the microscopic symmetries
of a material.   Since we have assumed that the material is at rest relative to some decomposition of Minkowski space as $\R\times \R^3$,
the relevant symmetries are the ones that preserve this decomposition, namely the product of the symmetries of the two factors.
We have already assumed space and time translation invariance in writing eqn. (\ref{tellme}) with no explicit dependence of $f$ on $\vec x$ or $t$.
The remaining symmetries to consider are the group $O(3)$ of rotations and spatial reflections of $\R^3$, along with time-reversal, $t\to -t$.

A homogeneous, isotropic substance like a liquid or a glass would usually be invariant under $O(3)$ and under time-reversal.
That  puts strong restrictions on the coefficients in the effective action, as we will discuss shortly in examples.
By contrast, in a crystal, the arrangement of the crystal axes will break $O(3)$ to a finite subgroup (possibly trivial, but non-trivial
for most simple crystals).   Even experiments that are not sensitive to what is happening at the atomic scale can easily see  violation
of $O(3)$ symmetry (and/or  time-reversal symmetry) because, as we will see in examples, this has a great bearing on what terms are present
in the effective action and what macroscopic phenomena follow.

Let us practice by making a Taylor series expansion of the effective action and discussing the significance of the first few terms.   We can start by adding to the Maxwell action 
a term linear in $\vec E$:
\be\label{ferro}I_\eff=\frac{1}{2}\int_{\R^{3,1}}\d t\,\d^3x \left(\vec E^2-\vec B^2\right)+\int_S\d t\, \d^3x \,\vec c\cdot \vec E.\ee
Here $\vec c$ must be a vector in $\R^3$ that is somehow characteristic of the material, so this term in the effective action will
not arise if a material has too much rotational symmetry.   A material with this term in the effective action is ``ferroelectric,'' meaning
that in the absence of any externally applied electric field, it spontaneously produces an internal electric field.   To show this,
one has to solve the Euler-Lagrange equations coming from the combined action $I_\eff$ of eqn. (\ref{ferro}) and show that 
a solution in which $\vec E $ and $\vec B$ vanish far away from the material has $\vec E$ nonzero inside it.  

Similarly, we can contemplate a material in which the effective action has a term linear in $\vec B$:
\be\label{nerro} I_\eff= \frac{1}{2}\int_{\R^{3,1}}\d t\,\d^3x \left(\vec E^2-\vec B^2\right)+\int_S\d t\, \d^3x \,\vec c'\cdot \vec B.\ee
In addition to being absent in a material with too much rotational symmetry, this term is absent in a material
that  possesses time-reversal symmetry ($\vec E$ and $\vec B$ are respectively even and odd
under time reversal).
As one finds by solving the Euler-Lagrange equations, a material in which this term is present in the effective
action is a ferromagnet, that is, it spontaneously generates an internal magnetic field.   Ferromagnetism
is familiar in everyday life (ordinary bar magnets are ferromagnets), but the most familiar ferromagnets do not really fit our discussion  since they are not
insulators.   However, there definitely are ferromagnetic (and ferroelectric) insulators.   A material can be both ferroelectric and ferromagnetic if the coefficients
$\vec c$ and $\vec c'$ are both nonzero. 

If a material has too much rotational symmetry to allow in the effective action a term linear in $\vec E$ or $\vec B$, it can still have a quadratic
term.   If we assume full rotational symmetry, for example,  the possible quadratic terms take a simple form\footnote{The coefficient $\gamma$
vanishes if time-reversal symmetry is present.}
\be\label{zolt}\int_S \d t\,\d^3 x \left(\epsilon \vec E^2+\mu \vec B^2+\gamma \vec E\cdot \vec B\right).  \ee
A material in which these are the lowest order contributions to the effective action is called a dielectric.
A  typical example is a piece of glass, or a transparent fluid such as water or air.
In a dielectric, the speed of light
is less than it is in vacuum -- as we discover if we add the corrections (\ref{zolt}) to the original Maxwell action and solve
the resulting Euler-Lagrange equations.   Another characteristic property of a dielectric, also implied by the same
Euler-Lagrange equations,  is that light is ``refracted,'' or bent at an angle (and also partly reflected), 
in passing between vacuum and a dielectric (or between two dielectrics with different values of $\epsilon$ and $\mu$, such as any two of air, water, and glass).  
Refraction of light in entering or leaving a piece of glass makes possible the construction of lenses.   

The effective action in general depends on derivatives of $\vec E$ and $\vec B$, so for example a possible term is
\be\label{olt} I_\eff=\cdots +\int_S \d t\d^3 x \sum_{i=1}^3\frac{\partial}{\partial x^i}\vec E\cdot \frac{\partial}{\partial x^i}\vec E. \ee
In the presence of such a term, the velocity at which light propagates through a dielectric, and also the angle at which it is refracted
in entering or leaving the dielectric, depends on its frequency or color.   This was observed by Descartes, who used a prism to separate
white light into all the colors of the rainbow.

In the case of a material with less symmetry, the quadratic terms in the effective action can be more complicated than was assumed
in eqn. (\ref{nerro}).  In  general one can have in the effective action
an arbitrary quadratic polynomial in the components of $\vec E$ and $\vec B$.   
One then runs into the phenomenon of birefringence.   The velocity of light inside a birefringent material
 depends on its direction of propagation and its polarization state, and a  light ray entering such a material from outside will
split into two rays, with different polarization states.  All this comes again by solving Maxwell's equations with a general quadratic
term in $\vec E$ and $\vec B$ included in the effective action.

Finally, we can consider the possible presence in the effective action of the integral of a term that is cubic or higher order in $\vec E$ and $\vec B$.
(A cubic term is not $O(3)$ invariant, so such a term is only possible in a material that does not have too much symmetry. A quartic term is possible
in any material.)    Once one
includes such terms in the effective action,  Maxwell's equations become nonlinear.
We enter the world of nonlinear optics, which has become important in the last few decades, with many strange and unfamiliar effects that are accessible
in very strong $\vec E$ and $\vec B$ fields.

What we have hopefully learned is that: (1) a material is characterized by an effective action; (2) the
effective action for a suitably chosen material can be more or less anything subject to general principles;
(3) a lot of physics follows from straightforward analysis of the effective action.   By considering only insulators, we kept things
simple.   This enabled us to consider effective actions for a $U(1)$ gauge field only -- the gauge field $A$ of electromagnetism 
-- without discussing the charges and currents inside the material.

We will need one more lesson about the effective action.   A full description of nature would involve the $U(1)$ gauge field $A$ and other
variables such as electrons and atomic nuclei.  Let us just write $\Phi$ for those variables.   So the microscopic action would be something like
\be\label{micro}I_\micro=I_\Maxwell+\tilde I(\Phi,A)=\frac{1}{4e^2}\int_{\R^{3,1}} F\wedge \star F+\tilde I(\Phi,A), \ee
where $\tilde I(\Phi,A)$ is the part of the action that describes the additional variables $\Phi$ and their interaction with the $U(1)$
gauge field $A$.  
The Euler-Lagrange equations for $A$ then become
\be\label{turnox} \d\star F +e^2\frac{\delta \tilde I(\Phi,A)}{\delta A}=0.\ee
The extra term that has appeared in Maxwell's equations is called the electromagnetic current:
\be\label{burnox}\J =e^2\frac{\delta I(\Phi,A)}{\delta A}. \ee
To be more precise, if we expand
\be\label{expo}\star \J =\rho \d t + \vec J\cdot \d\vec x,\ee
then $\rho$ is the electric charge density and $\vec J$ is the electric current.

In this article, we are assuming that the material of interest is an insulator, which has no relevant degrees of freedom.
What this means, roughly, is that for purposes of describing observations made at distances large compared to the atomic
scale, we can extremize $I_\micro$ with respect to $\Phi$ to get an effective action that depends on $A$ only:
\be\label{zeno} I_\eff=I_\Maxwell+\tilde I(A). \ee
This does not change the derivation of the electromagnetic current, which is now a function of $A$ only:
\be\label{zeddo}  \J= e^2\frac{\delta \tilde I(A)}{\delta A}.\ee
This formula gives us a framework to understand the currents and charges in an insulating material.  
For example, this formula can be fruitfully applied to get a fuller understanding of the ferroelectric, ferromagnetic,
and dielectric materials that we have encountered so far.  

Instead, we will turn in another direction, which will lead us back to the Chern-Simons function of a connection.

\section{Two-Dimensional Materials}

We are going to consider two-dimensional materials -- thin films.   
A thin film could be  a monoatomic layer floating in empty space, a very thin layer of one substance on the surface
of another, or in general anything that has two large spatial dimensions and one small one, and whose
small dimension is smaller than whatever distances will be probed by the experiments that we intend to discuss.
In effective field theory, we can  idealize this by thinking in terms of a purely two-dimensional material, infinitely thin in the third dimension.

A material with two space dimensions, when
we take into account time as a third dimension, occupies a three-dimensional submanifold $W$ of spacetime.
Actually, in discussing thin films, we do not want to assume that spacetime is simply Minkowski space.  Restricting to Minkowski space
would interfere with the topological argument that we will make presently (because the assumption that $W$ is embedded in $\R^{1,3}$,
throughout which the gauge field $A$ is defined, would give a way to resolve the indeterminacy of the Chern-Simons function).\footnote{Instead of generalizing from Minkowski
space to a more general four-manifold, one can make the following  argument in Minkowski space but using the fact that the relevant equations of physics make sense
in the presence of magnetic monopoles.   However, that would not lead to a shorter explanation.}
Thus, we will use a lesson that comes from general relativity:
 the equations that describe ordinary matter -- in the most complete known form, the standard model of particle physics -- make sense on an
arbitrary pseudo-Riemannian four-manifold $M$ which is endowed with a spin structure.   A spin structure is needed because, for example, electrons have spin $\frac{1}{2}$ and the Dirac
equation describing an electron  can only be defined  on a spin manifold.  It is of course no coincidence that the geometric structure that makes the Dirac equation
possible is called a  ``spin structure.'' 

Since ordinary matter is nonrelativistic to a very good approximation, we do not really need to consider an arbitrary four-dimensional spin manifold $M$.   It suffices
to consider the case that 
\be\label{yug}M=\R\times Y,\ee
where $Y$ is a three-dimensional spin manifold, and the metric of $M$ has the time-independent form $-\d t^2+g_Y$, for some time-independent metric $g_Y$ on $Y$.
We assume that the material under study is at rest, so its world-volume is of the form $W=\R\times D$, for some $D\subset Y$.   We write $g_D$ for the metric of $D$
and $\d^2x\sqrt{g_D}$ for its Riemannian measure.    We will consider a material that does not have spatial reflection symmetry, so the two-manifold $D$ must be oriented.
$D$ may be a compact two-manifold without boundary, or it may have a boundary.  The spin structure of $M$ induces one on $W$ and on $D$.

We again assume that the material under study is an insulator, so as before, we can describe it by an effective action for the gauge field $A$ only.   This
effective action can have the sort of terms with which we are by now familiar,
\be\label{zonot} I_\eff =\cdots +\int_W\d t \d^2x \sqrt{g_D} \left( \vec c\cdot \vec E+\vec c'\cdot \vec B+\epsilon \vec E\cdot \vec E+\cdots\right),\ee
with the sole difference that now the integral is taken over a submanifold $W$ of spacetime that has dimension $2+1$ rather than
$3+1$.
All of these terms are expressed in terms of $\vec E$ and $\vec B$ only, and lead to the sort of phenomena that we have already discussed
for three-dimensional materials.   There is only one
possible term in the effective action that cannot be written just in terms of $\vec E$ and $\vec B$, and will lead to something that is qualitatively new and special to two-dimensional
materials. This is the Chern-Simons function.

The definition of the Chern-Simons function for a $U(1)$ gauge field on a three-manifold $W$
involves a few subtleties that go beyond what we have described earlier.   We will discuss the definition assuming that $W$ is a closed three-manifold.
In the physical application, it is important to take into account boundaries of $W$, but we postpone that discussion.

A $U(1)$ gauge field $A$ on a manifold $W$ can be viewed as a unitary connection on a complex line bundle $\L\to W$.  A complex line bundle $\L$
has, of course, no second Chern class, but it does have a first Chern class $c_1(\L)$, which at the level of differential forms is represented
by $F/2\pi$, where $F=\d A$ is the curvature or electromagnetic field strength.  As an  integer-valued characteristic class of degree 4, 
we can consider $c_1(\L)^2$.  Hence if  $X$ is a closed four-manifold and  $\L\to X$ is a complex line bundle with connection $A$ and curvature $F$, the quantity 
\be\label{pelf} \int_X c_1(\L)^2=\int_X\frac{F\wedge F}{4\pi^2} \ee
is an integer. But  if $X$ is a spin manifold, then the intersection form on $H^2(X;\Z)$ is even, and accordingly $\int_X c_1(\L)^2$ is an even integer.
In that situation, then, the basic integer-valued invariant is actually one-half the integral of $c_1(\L)^2$ or 
\be\label{nelf} \int_X \frac{F\wedge F}{8\pi^2}. \ee

If  $W$ is  a three-dimensional spin manifold  endowed with a complex line bundle $\L$ with connection $A$, there always exists an oriented  four-manifold $X$ with
boundary $W$ such that the spin structure of $W$ and the line bundle $\L$ and connection $A$ all extend over $X$.
We 
define $\CS_X(A)$ by the same integral as in eqn. (\ref{nelf}), now over the manifold $X$ with boundary:
\be\label{gome} \CS_X(A)= \int_X \frac{F\wedge F}{8\pi^2}. \ee
The same argument as before shows that, mod $\Z$,  $\CS_X(A)$ depends only on the restriction of $\L$ and $A$ to $W$.   So we can forget about $X$ and 
define $\CS(A)$ as an invariant, valued in $\R/\Z$,
of the $U(1)$ connection $A$ over a three-dimensional spin manifold $W$.   In general, $\CS(A)$ does depend on the spin structure of $W$.
For an example, let $W$ be a three-torus, factored as ${\Bbb T}^3={\Bbb T^2}\times S^1$, and let $A$ be the pullback from ${\Bbb T}^2$ of a connection on a degree
1 line bundle over ${\Bbb T}^2$.   Then $\CS(A)$ equals 0 or $1/2$ depending on the spin structure of ${\Bbb T}^3$.

The following fact is often useful in practice.   If the complex line bundle $\L\to W$ is trivial and we pick a trivialization, then the connection $A$ becomes an ordinary one-form,
and the definition of $\CS(A)$ reduces to
\be\label{zimbo} \CS(A)=\frac{1}{8\pi^2}\int_W A\wedge \d A. \ee  Though this is not a completely general formula for $\CS(A)$, it suffices for purely local computations
(in which the global triviality or non-triviality of $\L$ is not important).
For example,  we can use this formula to directly verify the invariance of $\CS(A)$ under gauge transformations that are homotopic to the identity.   Such a gauge transformation acts on $A$
by $A\to A+\d \phi$, for some real-valued function $\phi$ on $W$. By integration by parts in the preceding formula, one can verify the  invariance of $\CS(A)$ under such gauge
transformations.    As another example, we will
do an elementary computation that we will need several times in what follows.   Let us compute
the variation of $\CS(A)$ under a general first order variation of $A$, $A\to A+\delta A$.  We have, after integrating by parts and using $F=\d A$,
\be\label{mimbo}\CS(A) \to \CS(A)+\frac{1}{8\pi^2}\int_W( \delta A\wedge \d A+A\wedge \d \delta A)=\CS(A) +\frac{1}{4\pi^2}\int \delta A\wedge F.\ee
Thus, a critical point of the functional $\CS(A)$ is precisely a connection with $F=0$, that is, a flat connection.  

We now ask the following question.   Given that $\CS(A)$ is only defined mod $\Z$, is it possible for a multiple of $\CS(A)$ to appear as a contribution to the effective action?
In classical physics, we would answer this question as follows.   As we noted earlier, the sole importance 
of the action in classical physics is that the equations of motion -- the Euler-Lagrange equations --
say that the action is stationary under small variations.  That condition is invariant under adding a constant to the action.  So in classical physics, it makes perfect sense to have
an action that is only well-defined modulo an additive constant.   

Matters are different in quantum mechanics.   According to Feynman, a quantum system can be described by integrating over all possible classical orbits, with each
orbit being weighted by a factor $e^{\i I/\hbar}$, where $I$ is the action for the orbit and 
$\hbar$ is Planck's constant.   Therefore, the action $I$ does not have to be well-defined as a real-valued function, but it does have to be well-defined mod
$2\pi \hbar \Z$.   This was first explained by Dirac in his analysis of magnetic monopoles \cite{Dirac}.  It is convenient, however, and quite common among physicists,
to work in units with $\hbar=1$.  We have actually already done so in omitting factors of $\hbar $ in some previous formulas, and we will do so going forward.
With practice, the appropriate power of $\hbar$ multiplying any quantity can easily be restored on dimensional grounds.  

Since $\CS(A)$ is defined mod $\Z$, and the action must be well-defined mod $2\pi\Z$, we conclude that a multiple of $\CS(A)$ can indeed be present in the effective action of a two-dimensional material,
but the coefficient multiplying it must be of the form $2\pi  k$ for some integer $k$:
\be\label{cseff}I_\eff=\cdots +2\pi  k \CS(A). \ee   Integrality of $k$ will ensure that $e^{\i I_\eff}$ is well-defined.   
Thus, two-dimensional insulating materials are classified by an integer invariant $k$.  

The existence of this integer invariant of a two-dimensional insulating material was discovered quite unexpectedly in the laboratory more than 40 years ago \cite{Klitzing}.
How was the experimental discovery made?  What experimentally accessible observable depends on the integer $k$?
To answer this question, we will specialize to a realistic situation.   We take spacetime to be Minkowski space $\R^{1,3}$, parametrized
as usual by $t $ along with  $\vec x=(x,y,z)$.   We take the material of interest to occupy a region of the $x-y$ plane at $z=0$.   

The definition of an electrical conductor is that,  in the presence of a weak, constant
 electric field in, say, the $x$ direction, with $F=E_x \d t \d x$, there is an induced electric current in the $x$ direction that
is proportional to $E_x$.  
The constant of proportionality between the current and the electric field is called the
ordinary or longitudinal conductivity.  No term in the effective action (\ref{zonot}) or (\ref{olt}) will 
lead to such an effect.   A short computation shows that the contribution of any such term to the current as defined in eqn. (\ref{zeddo}) is proportional
to derivatives of the electromagnetic field, not to the electromagnetic field itself.  
So the ordinary  conductivity  vanishes in any material in which the effective action can be described as a function  of $A$ only, 
without including degrees of freedom intrinsic to the material.  That is why we have referred to such materials as insulators.

  However, there is something more subtle that can happen: an electric field in the $x$ direction may produce a current
in a perpendicular direction.     This effect is called Hall conductivity; a version of it was discovered in the nineteenth century, in a context in which
quantum effects were quite unimportant \cite{Hall}.   In the case of a two-dimensional material that is localized in the $x-y$ plane, the only possible
direction within the material and orthogonal to the $x$ direction is the $y$ direction.   So Hall conductivity will mean that an electric field in the $x$ direction produces a current in the
$y$ direction.

We will now show that this actually happens in a  two-dimensional material with $k\not=0$.   For this, we just 
expand the electric current as $\vec J=J_x \d x + J_y \d y + J_z \d z$ and from eqns. (\ref{zeddo}) and (\ref{expo}), we find that the current in the $y$ direction due to
the presence of the Chern-Simons term in the effective action is
\be\label{homerun} J_y=e^2\frac{\delta}{\delta A_y} 2\pi k \CS(A). \ee  
Here the definition of $\frac{\delta}{\delta A_y} 2\pi k \CS(A)$ is that 
 the first order change in $\CS(A)$ if $A$ is varied by $A\to A+\delta A$ with $\delta A=\d y \delta A_y$ is $\int_W \delta A_y \frac{\delta}{\delta A_y} 2\pi k \CS(A)$.
 Using eqn. (\ref{mimbo}) and 
 fixing the orientation of $W$ by the three-form $\d t\, \d x\, \d y$, we then compute
\be\label{omerun}J_y =\frac{e^2 k}{2\pi} E_x  \delta(z). \ee
The delta function $\delta(z)$ is just telling us that the current is supported at $z=0$, that is inside the material.   (Of course, in a more realistic model of a thin but
not infinitely thin material, the delta function would be replaced by a ``bump function'' supported near $z=0$ and integrating to 1.)

Thus we have arrived at the quantum Hall effect, with the constant of proportionality between $E_x$ and $J_y$ being $e^2 k/2\pi$.  Of course, we have computed in units with
$\hbar=1$.  Restoring $\hbar$ in the formula,
the Hall conductivity is
\be\label{dico}\sigma_{xy}=\frac{ke^2}{2\pi \hbar}.\ee
  Thus, a two-dimensional insulator can exhibit a Hall conductivity with a coefficient that will be an integer multiple of $\frac{e^2}{2\pi\hbar}$.

As already noted, this effect was rather dramatically discovered\footnote{\label{showme} In more detail, a two-dimensional material was studied in the presence of a strong magnetic field perpendicular to the plane of the material.
In certain ranges of the magnetic field strength, the material is an insulator, and $k$ is found to be an integer.  As the magnetic field is increased, the material ceases
to be an insulator; with further increase of the magnetic field, it may become an insulator again, typically 
with a smaller  integer value of $k$.  In some modern experiments, integer
Hall conductivity is observed with no applied magnetic field, as first suggested in \cite{Haldane}. See  \cite{Xiaoliang} for a review.} in 1980 in advance of any theoretical prediction \cite{Klitzing}.   In real samples, $k$ is found to be a small
integer
 and its integrality is measured with incredible precision, of order one part in $10^9$, making the Hall conductivity one of the most precisely measured of all physical
observables.  In fact, tests of the quantization of the Hall conductivity have been so precise as to strain the independent knowledge of the ratio $e^2/\hbar$.
At times, these measurements have given the most precise measurement of $e^2/\hbar$ (given the prediction that $k$ is going to be an integer).

In real samples, both even and odd values of $k$ are observed.   In the context of the derivation given here, this shows that it was
indeed important to take into account the fact that spacetime is endowed with a spin structure.   Had we not taken this into account, we would have predicted that $k$
must be an even integer.

In defining $\CS(A)$, we  treated $W$ as a closed three-manifold. This was necessary, because the definition of $\CS(A)$ 
involved viewing $W$ as the boundary of a compact four-manifold, which  of course is only possible if $W$ is a closed manifold.
  In our application, the assumption that $W$ is closed is not realistic, since $W=\R\times D$, where $\R$ is parametrized by the time $t$ and $D$
is an oriented two-manifold with or without boundary.   Even if $D$ has no boundary, $W$ is not a closed manifold, because $\R$ is not compact.  However, the  ``ends'' of $W$ at
$t\to \pm \infty$ do not actually interfere with what we have said in an essential way.   
Roughly,  we can restrict ourselves to experiments in which the connection $A$
vanishes outside a very large time interval $-T\leq t\leq T$.   In that case, we can compactify the ends of $W$, with $A$ extended by 0 over the compactification,
 and reduce to the case that $W$ is
a closed manifold.    (With a little more care, one can also describe what happens if $A$ does not vanish for $t\to\pm \infty$.)

But the picture really does change if $D$ has nonempty boundary $\partial D$.   Here I will give
only a bare sketch of a rather elaborate story.    We have characterized an insulator as a material that has no relevant degrees of freedom,
so that its effective action can be defined as a function only of the $U(1)$ gauge field $A$ of electromagnetism.  However, in the case of a two-dimensional material with a nonzero
Hall conductivity, this assumption, while valid if $D$ is a closed Riemann surface, fails along  $\partial D$.   Rather, there are relevant degrees of freedom that propagate only on $\partial D$, leading to what
are usually called ``edge currents'' in a quantum Hall material.   A slightly abstract description is as follows.  Let $W$ be an oriented three-manifold with boundary $\Sigma$.  
For $A$ a unitary connection on a complex line bundle $\L\to W$, one cannot define $\exp(2\pi \i k \CS(A))$ as a complex number.   However, one can define it as a
vector of unit norm in a complex line $\RR$ that depends on the restriction of $A$ to $\Sigma$ \cite{Singer}.   To compensate for this, one can assume on $\Sigma$ the presence
of a two-dimensional quantum field theory whose partition function or Feynman path integral $Z$ can likewise not be defined as a complex number, but is well-defined
as an element of $\RR^{-1}$.  Then the product $Z \exp(2\pi\i k \CS(A))$ is a well-defined complex number, though neither factor is separately.  So the combination of the quantum
Hall system with a boundary theory of partition function $Z$ is well-defined.   Physicists describe this
by saying that the boundary quantum field theory  in question is ``anomalous,''  and its anomaly is compensated via ``anomaly inflow'' \cite{CallanHarvey}
 by the ill-definedness of $\exp(2\pi\i \CS(A))$ as a complex number.  In the context of a quantum Hall system with a given value of $k$, the
anomalous quantum field theory that appears on  $\Sigma$ is not uniquely determined; different materials with the same value of $k$ may have different boundary
theories.   The simplest possibility is a theory of ``free fermions'' in which $Z$ is the $k^{th}$ power of the determinant of a certain Dirac operator.

\section{The Fractional Quantum Hall Effect}\label{four}

Physicists had scarcely come to understand why the coefficient $k$ in the Hall conductivity of a two-dimensional insulator is always an integer
when it was discovered that for some materials this is not so.   A quantum Hall effect was observed with $k=1/3$ \cite{Tsui}.   
Since then, quantum Hall behavior has been observed with a variety of rational values of $k$,   always with a relatively small denominator.  This phenomenon is known 
as the fractional quantum Hall effect.\footnote{Like the integer quantum Hall effect, the fractional one was first observed in the presence of a magnetic field; the remarks of footnote \ref{showme} are again applicable.   Quite recently, the fractional quantum Hall effect was observed without an applied magnetic field
\cite{FAQHE}.}

What is going on?   The original and still very illuminating explanation involved a many electron wavefunction with remarkable properties \cite{Laughlin2}, but
 an alternative explanation\footnote{On the complex history of this approach, see the historical remarks in chapter 4 of \cite{Stone}.  Important early papers include
 \cite{Zhang,Read} as well as other articles reprinted in \cite{Stone}.}  that is in the spirit of the present article  relies on a topological field theory  based on the Chern-Simons function.

We characterized an insulator as a material with no relevant degrees of freedom, so that we can describe its interaction with the electromagnetic gauge field $A$
by an effective action that is a function of $A$ only.    To understand the fractional quantum Hall effect, we have to relax this assumption and
say only that any relevant degrees of freedom of the material are trivial locally and any information they store is of a global nature.   Differently put, any relevant
degrees of freedom in the material are topological in nature and describe a topological field theory.  

For a particularly simple example, consider  a two-dimensional material in which the only relevant degree of freedom is a $U(1)$ gauge field $a$, which is a connection on a complex line bundle
$\M$ over the worldvolume $W$ of the material.      Unlike the gauge field $A$ of electromagnetism,
which exists throughout spacetime, $a$ is an ``emergent'' gauge field that exists only in the worldvolume of the material.   The effective action for $a$ (ignoring its coupling to $A$) 
is assumed to be a multiple of the Chern-Simons function:
\be\label{toldo} I_\eff(a) = 2\pi r \CS(a). \ee
Of course, $r$ has to be an integer.   Other terms may be present in the effective action, such as a ``Maxwell'' term $\int_W f\wedge \star f$, where $f=\d a$ is the curvature of $a$.
However, as long as $r\not=0$, the Chern-Simons term is the dominant one at long distances and we can ignore other terms.

The magic of the Chern-Simons action
is that, as the function $\CS(a)$ for a gauge field $a$ on a three-manifold $W$ is defined with no structure required on $W$ except an orientation (and a spin structure if $r$ is odd), a theory whose action 
is a multiple of $\CS(a)$ is going to be a topological field theory, not depending on any Riemannian or pseudo-Riemannian metric on $W$.   Such a theory is
only going to capture global information.

We can readily verify this at the classical level.  As discussed in relation to eqn. (\ref{mimbo}), the Euler-Lagrange equations for a critical point of the function $\CS(a)$
is $f=0$, in other words a solution of the Euler-Lagrange equations is a flat connection.   But a flat connection on a manifold $W$ 
is trivial locally.   In particular, all classical solutions look the same locally; they can only be distinguished by global holonomies.

Physicists refer to this theory -- with an abelian gauge group, in this case $U(1)$, and an action that is a multiple of the Chern-Simons function -- as ``abelian Chern-Simons
theory.''   Nonabelian Chern-Simons theory is then a gauge theory with a nonabelian gauge group and an action that is a multiple of the Chern-Simons function.  A variant of 
abelian Chern-Simons theory was first studied in \cite{Schwarz}.   Nonabelian Chern-Simons theory was introduced in \cite{Witten}.

The reader may well be wondering ``where did the field $a$ come from?''  Here we meet the wonders of many-body quantum physics.   Whatever material it is that
we are studying, it is made microscopically from electrons and atomic nuclei.   They are described by the Schr\"{o}dinger equation (or the more complete standard model
of particle physics).  A macroscopic piece of matter contains a huge number of electrons and atomic nuclei -- roughly $10^{23}$ per gram.   A vast number of quantum particles
 working together can generate an incredible variety of surprising phenomena.\footnote{One of these, namely superconductivity, was described in
 my previous article in this journal \cite{WittenB}.}   Among those surprises, apparently, is the generation of the field $a$ (and various
 generalizations thereof).   Microscopically, the material is made from electrons and atomic nuclei, and there is no $a$ field.   Macroscopically, at distances very large
 compared to the atomic scale, an effective description via the field $a$ spontaneously emerges.   The physics literature contains various heuristic explanations of how
 this happens, but ultimately the statement is more a theoretical interpretation of experimental facts than a theorem about the solutions of the Schr\"{o}dinger equation.
 
 At any rate, once one assumes emergence of the field $a$, one can make a model of the fractional quantum Hall effect.   In addition to $\CS(A)$ and $\CS(a)$,
 we have to consider one more interaction that couples the two gauge fields $A$ and $a$.    We start with the four-dimensional characteristic class $c_1(\L)c_1(\M)$, represented in terms
 of differential forms by the integer-valued invariant $\int_X \frac{F\wedge f}{4\pi^2}$ on a closed oriented four-manifold $X$.  Note that even if a spin structure is assumed, this can
 be an arbitrary integer, not necessarily even, so spin will not play a role.
  If  $X$ is an oriented four-manifold with  boundary $W$,
 the same integral defines what we will call
 \be\label{inco}\CS_X(A,a)=\int_X \frac{F\wedge f}{4\pi^2}.\ee
 The usual argument shows that, mod $\Z$, $\CS_X(A,a)$ only depends on the gauge fields $A$ and $a$ restricted to $W=\partial X$, and not on how they are extended over $X$.
 So we can forget $X$ and define $\CS(A,a)$ as an invariant valued in $\R/\Z$.   It therefore may again appear in the effective action with a coefficient $2\pi s$, $s\in \Z$.
 
The effective action that describes the material -- meaning the field $a$, which we assume is the only relevant degree of freedom in the material --
 and the interaction with the electromagnetic gauge field $A$ is then
  \be\label{winco}I_\intt(A,a)= 2\pi k \CS(A) + 2\pi s \CS(A,a) + 2\pi r \CS(a) ,\ee with integers $k,s,r$.
 If the two line bundles $\L$ and $\M$ are trivial and we pick  trivializations so that $A$ and $a$ become ordinary one-forms, this action can be written
 \be\label{zinco}I_\intt(A,a)=\int_W\left( \frac{k}{4\pi} A\wedge \d A +\frac{s}{2\pi} A\wedge \d a +\frac{r}{4\pi} a\wedge \d a\right),\ee
 generalizing eqn. (\ref{zimbo}).   
Generalizing the computation in eqn. (\ref{mimbo}), the Euler-Lagrange equation for $a$ is
\be\label{eom} sF+r f=0.\ee
So $f$ is not zero in the presence of an electromagnetic field, but it is uniquely determined by the electromagnetic field and therefore the field $a$ locally
contains no information not present in the electromagnetic field.

Now we can return to the case of a material localized in the plane $z=0$ in Minkowski space
and compute the Hall conductivity.  
We simply use the formula (\ref{homerun}), but with $2\pi k \CS(A)$ replaced by $I_\intt(A,a)$.   We get
\be\label{ivo}J_y=\frac{e^2}{2\pi}(k E_x+s e_x), \ee
where we have expanded
\be\label{nivo}f = \d t \,\d x \,e_x+\d t\,\d y\, e_y + \d x \d y\, b. \ee
But the equation of motion (\ref{eom}) gives $f=-\frac{s}{r} F,$
so the current is
\be\label{wivo} J_y=\frac{e^2}{2\pi}\left(k-\frac{s^2}{r}\right)E_x.\ee
Thus  the coefficient of the Hall conductivity is the rational number $k-s^2/r$, and we have arrived at a model of the fractional quantum Hall effect.  For example,
if $k=0$, $s=1$, $r=-3$, we get a model in which the coefficient is $1/3$, as in the original experimental discovery.

From here, one could continue  in many different directions.  The role of topological field theory in modeling the fractional quantum Hall
effect has helped inspire  many further ideas about topological field theories in condensed matter physics, or what are sometimes
called topological phases of matter \cite{Sachdev,FH,Frohlich}.  There is a dream of one day using a topological phase of matter to build a quantum computer \cite{Kitaev}.  Almost
all known (unitary) topological field theories in three spacetime dimensions are Chern-Simons theories \cite{MooreSeiberg}.
 The model of the fractional quantum Hall effect that we have discussed based on abelian Chern-Simons theory is
actually believed to be a good model of most such systems that are experimentally observed, including the original ones.  But there is much more to say about this
model, among other things concerning the spin and electric charge of localized ``quasiparticle'' excitations \cite{Halperin}, and the ground state degeneracy when the
theory is formulated on a Riemann surface of genus $g>0$ \cite{Witten,Wen}.    In addition, there are a few observed fractional quantum
Hall systems, notably one with $k=5/2$ \cite{FiveHalves}, which are believed to require a description by a nonabelian Chern-Simons theory, as pioneered  in
\cite{MooreRead}. Decisive experimental proof of this remains elusive.   Nonabelian Chern-Simons gauge theory is also a very rich subject from a mathematical
point of view, as it offers a framework \cite{Witten} to understand the Jones polynomial of a knot and more general quantum invariants of knots and three-manifolds
in terms of quantum field theory.   

\vskip1cm
 \noindent {\it {Acknowledgements}}  
  Research supported in part by NSF Grant PHY-2514611.
 \bibliographystyle{unsrt}

\begin{thebibliography}{99}


\bibitem{BPST}A. A. Belavin, A. M. Polyakov, A. S. Schwartz, and Yu. S. Tyupkin, ``Pseudoparticle Solutions of the Yang-Mills Equations,''
Phys. Lett. {\bf B59} (1975) 85-87.

\bibitem{THooft}G. 't Hooft, ``Symmetry Breaking Through Bell-Jackiw Anomalies,'' Phys. Rev. Lett. {\bf 37} (1976) 8-11.

\bibitem{Donaldson}S. Donaldson, ``An Application of Gauge Theory To Four-Dimensional Topology,''   J. Diff. Geom. {\bf 18} (1983) 279-315.

\bibitem{Dirac}P. A. M.  Dirac, ``Quantised Singularities in the Electromagnetic Field,'' Proc. Roy. Soc. {\bf A133} (1931) 60-72.

\bibitem{Klitzing}
K. v. Klitzing,  G. Dorda, and  M. Pepper,  ``New Method for High-Accuracy Determination of the Fine-Structure Constant Based on Quantized Hall Resistance,''
Phys. Rev. Lett. {\bf 45} (1980) 494-497.

\bibitem{Laughlin}R. B. Laughlin, ``Quantized Hall Conductivity In Two Dimensions,'' Phys. Rev.{\bf  B23} (1981)  5632-5633.

\bibitem{Hall}   E. Hall, ``On a New Action of the Magnet on Electric Currents,''  American  Journal of Mathematics {\bf 2} (1879) 287-292.


\bibitem{Haldane}
F. D. M. Haldane, ``Model For A Quantum Hall Effect Without Landau Levels,''  Phys. Rev. Lett. {\bf 61}  (1988) 2015-8.

\bibitem{Xiaoliang}
Chao-Xing Liu, Shou-Cheng Zhang, and Xiao-Liang Qi, ``The Quantum Anomalous Hall Effect,'' arXiv:1508.07106.


\bibitem{Singer}
T.R. Ramadas, I.M. Singer, and J. Weitsman, ``Some Comments on Chern-Simons Gauge Theory,''
 Commun. Math. Phys. {\bf 126} (1989) 409-420.

\bibitem{CallanHarvey}
C. G. Callan, Jr., and J. A. Harvey, ``Anomalies And Fermion Zero-Modes On Strings And Domain
Walls,''  Nucl. Phys. {\bf B250} (1985) 427-36.

\bibitem{Tsui}
D.C. Tsui,  H.L. Stormer, and  A.C. Gossard, ``Two-Dimensional Magnetotransport in the Extreme Quantum Limit,'' Phys. Rev. Lett.  {\bf 48} (1982)  1559-1562.

\bibitem{FAQHE}
H. Park, J,  Cai, E. Anderson, Y.  Zhang, J.  Zhu, X. Liu, C. Wang, W. Holtzmann, C. Hu, Z. Liu, T.  Taniguchi, K. Watanabe, J-h. Chu, T. Cao, L. Fu, W.  Yao, C-Z. Chang, D.
 Cobden, D. Xiao,  and X. Xu, ``Observation of Fractionally Quantized Anomalous Hall Effect,''  Nature {\bf 622} 74-9, arXiv:2308.02657.


\bibitem{Laughlin2}
R. B. Laughlin, ``Anomalous Quantum Hall Effect: An Incompressible Quantum Fluid with Fractionally Charged Excitations,'' Phys. Rev. Lett.  {\bf 50}  (1983) 1395-1398.

\bibitem{Stone}
M. Stone, {\it Quantum Hall Effect} (World Scientific, 1992).

\bibitem{Zhang}
S. C. Zhang, T. H. Hansson, and S. Kivelson, ``Effective Field Theory Model for the Fractional Quantum Hall Effect,''
Phys. Rev. Lett. {\bf 62} (1989) 82-5.

\bibitem{Read}
N. Read, ``Order Parameter and Ginzburg-Landau Theory for the Fractional Quantum Hall Effect,'' Phys. Rev. Lett. {\bf 62} (1989) 86-89.


\bibitem{Schwarz}
A. Schwarz, ``The Partition Function of a Degenerate Functional,'' Commun. Math. Phys. {\bf 67} (1979) 1-16.

\bibitem{Witten}
E. Witten, ``Quantum Field Theory and the Jones Polynomial,'' Commun. Math. Phys. {\bf 121} (1989) 351-399.

\bibitem{WittenB}
E. Witten, ``From Superconductivity and Four-Manifolds to Weak Interactions,'' Bull. AMS {\bf 44} (2007) 361-391.

\bibitem{Sachdev}
S. Sachdev, {\it Quantum Phases of Matter} (Cambridge University Press, 2023).

\bibitem{FH}
D. S. Freed and M. J. Hopkins, ``Reflection Positivity and Invertible Topological Phases,''  Geom. Topol. {\bf 25} (2021) 1165-1330.

\bibitem{Frohlich}
J. Fr\"ohlich, ``Gauge Invariance and Anomalies in Condensed Matter Physics,'' J. Math. Phys.  {\bf 64} (2023) 031903,
arXiv:2303.14741.

\bibitem{Kitaev}
A. Kitaev,  ``Fault-Tolerant Quantum Computation by Anyons,'' Ann.  Phys, {\bf 303} (2003) : 2-30.

\bibitem{MooreSeiberg}
G, W. Moore and N. Seiberg, ``Taming The Conformal Zoo,''  Phys. Lett, {\bf220B} (1989) 422-30.

\bibitem{Halperin}
D. E. Feldman and B. I.  Halperin, ``Fractional Charge and
Fractional Statistics in the Quantum Hall Effects,'' Rep. Prog.
Phys. {\bf 84}  (2021) 076501.

\bibitem{Wen}
X.-G. Wen, ``Vacuum Degeneracy of Chiral Spin States in Compactified Space,'' Phys. Rev. {\bf B40} (1989)  7387-90.


\bibitem{FiveHalves}
R. Willett, P.  J.Eisenstein, H. L. St\"{o}rmer, D. C. Tsui,
A. C. Gossard, J. H. English, ``Observation of
an Even-Denominator Quantum Number in the Fractional Quantum Hall Effect,''
Phys. Rev. Lett. {\bf 59} (1987)
1776-1779.

\bibitem{MooreRead}
G Moore and  N Read, ``Nonabelions in the Fractional Quantum Hall Effect,'' Nucl. Phys.  {\bf B360} (1991) 362-396.

\end{thebibliography}

\end{document}